\documentclass[preprint,prb,aps,showpacs,showkeys]{revtex4}
\usepackage{graphicx} 
\usepackage{dcolumn} 
\usepackage{bm} 
\usepackage{amsmath,amssymb}

\begin{document}

\title{Observation of Fano-Resonances in Single-Wall Carbon Nanotubes}

\author{B.\ Babi{\'c}}
\author{C. Sch{\"o}nenberger}
\email{Christian.Schoenenberger@unibas.ch}
\affiliation{Institut f\"ur Physik, Universit\"at Basel, Klingelbergstr.~82, CH-4056 Basel, Switzerland }
\date{\today}

\begin{abstract}
We have explored the low-temperature linear and non-linear
electrical conductance $G$ of metallic carbon nanotubes (CNTs),
which were grown by the chemical-vapor deposition method. The high
transparency of the contacts allows to study these two-terminal
devices in the high conductance regime. We observe the expected
four-fold shell pattern together with Kondo physics at
intermediate transparency \mbox{$G\alt 2e^2/h$} and a transition
to the open regime in which the maximum conductance is doubled and
bound by $G_{max}=4e^2/h$. In the high-$G$ regime, at the
transition from a quantum dot to a weak link, the CNT levels are
strongly broadened. Nonetheless, sharp resonances appear
superimposed on the background which varies slowly with gate
voltage. The resonances are identified by their lineshape as Fano
resonances. The origin of Fano resonances is discussed along the
modelling.
\end{abstract}
\maketitle
\section{Introduction}

The Fano resonance (FR) is a universal physical phenomenon which
has historically been discovered as asymmetric line profiles in
spectra of rare gases.\cite{Lane} This effect has successfully
been interpreted by U. Fano in terms of the interference between
an auto-ionized state and the continuum.\cite{Fano} FRs have for
example been observed in the spectroscopy of atoms and
molecules,\cite{Lane} in electron, neutron and Raman
scattering.\cite{Adair,Simpson,Cerdeira}

Generally speaking, the interference of a resonant state (the
resonant channel) with a continuum (the non-resonant channel)
gives rise to Fano line-shapes. An illustration is shown in
Fig.~\ref{FanoFig1}a. This phenomenon can also naturally arise in
coherent electrical transport through nanostructures. Indeed, Fano
line-shapes were observed in the differential electrical
conductance \mbox{$dI/dV$ vs. $V$}, while tunneling with a
scanning tunneling microscope through an impurity atom on a metal
surface.\cite{Madhavan,Schneider} The first observation of FRs in
mesoscopic devices has been reported by G\"ores \textit{et al.} in
a single-electron transistor fabricated into a gated
two-dimensional electron gas.\cite{Gores} Thereafter, FRs have
been found in several studies of charge transport through
similarly fabricated `two-dimensional' quantum dots, both for
single quantum dots and quantum dots embedded into an
Aharonov-Bohm
ring.\cite{Kobayashi-2002,Kobayashi-2003,Haug-2003,Kobayashi-more}
Recently, several groups have reported the observation of FRs in
multi-wall carbon nanotubes (MWNTs). Kim \emph{et al.} \cite{Kim}
observed the FR on crossed MWNTs, while Yi \emph{et al.} \cite{Yi}
reported on FRs measured in MWNT bundles. Furthermore, Fano
resonances have been measured on an individual
MWNT.\cite{Chandrasekhar} However, similar observations have not
yet been reported for single-wall carbon nanotubes (SWNTs).

Following the notion of U.~Fano,\cite{Fano} the energy-dependent
conductance $G(E)$ of a Fano resonance observed in a transport
measurement can be described in the following form:\cite{Gores}
\begin{equation}\label{Fano1}
  G(\epsilon)=G_{nonres} + G_{res}
  \frac{(\epsilon+q)^2}{\epsilon^2+1},
\end{equation}
where \mbox{$G_{nonres}$ and $G_{res}$} denote incoherent and
coherent contributions to the conductance.
The detuning of the energy $E$ from the center of the resonance $E_0$
is described by the dimensionless parameter \mbox{$\epsilon\equiv2(E-E_0)/\Gamma$},
where $\Gamma$ denotes the width of the resonant state.
\mbox{$q$} is the so-called asymmetry parameter. Its
magnitude is proportional to the ratio of
the transmission amplitudes of the resonant and non-resonant
channel. In the original Fano work,\cite{Fano1} $q$ was
introduced as a real parameter, in general however, it must be
treated as a complex quantity.\cite{Clerk}
In the limit \mbox{$q\rightarrow \infty$}, resonant transmission
dominates which leads to symmetric Breit-Wigner resonances, see Fig.~\ref{FanoFig1}b.
In the opposite limit \mbox{$q\rightarrow 0$}, the resonant transmission appears
as an anti-resonance, i.e. a symmetric dip.
In all other cases e.g. \mbox{$q=1$}, asymmetric line-shapes are obtained.
These asymmetric line-shapes are characteristic for the Fano effect and that is why
one refers to them as Fano line-shapes.

Since the Fano resonance is the result of an interference effect,
its line-shape is sensitive to the phase difference between the
two transmission pathways. Fano resonances can therefore provide
essential information on dephasing in mesoscopic
systems.\cite{Clerk,Kobayashi-2002,Kobayashi-2003,Haug-2003} In
this respect it has the same power as all other two-path
interference experiments, for example the Aharonov-Bohm
experiment.\cite{Yacoby}

Here we report on the observation of Fano resonances in electrical transport
measurements on SWNTs. In contrast to previous studies, which were focussed on
two-dimensional quantum dots, SWNTs can be classified as one-dimensional quantum dots.
MWNTs are different, because it has been shown that several subbands are in general
occupied,\cite{Krueger,Buitelaar} giving rise to the possibility that more than one
quantum dot state may participate in transport.

\section{Experimental}

Single wall carbon nanotubes (SWNTs) have been grown from patterned
catalyst islands by the chemical vapor deposition (CVD) method on
\mbox{Si/SiO$_2$} substrates.\cite{Babic2003} The degenerately doped silicon,
terminated by a \mbox{$400$\,nm} thick \mbox{SiO$_2$} layer, is used
as a back-gate to modulate the electrochemical potential of the
SWNT electrically contacted with a source and drain terminal.
Electrical contacts to individual tubes are patterned by electron-beam lithography.
Once the samples are made, semiconducting and metallic
SWNTs are distinguished by the dependence of their electrical
conductance $G$ on the gate voltage $V_g$ measured at room
temperature.\cite{BabicKirchberg2004}
Here, we focus on measurements performed on metallic SWNTs with
relatively low-ohmic contacts. We found independently of the work by Javey {\it et al.} \cite{Dai-Pd-2003}
that palladium makes excellent contacts to SWNTs. There is no need for an additional
post-growth treatment (e.g. annealing).\cite{BabicKirchberg2004,Avouris-Annealing}
Most of the samples show two-terminal conductances of \mbox{$ G > e^2/h$}
which is essential for the studies discussed below.

The electrical characterization of the devices has been
performed at \mbox{$300$\,mK} in a \mbox{$^3$He} system.
We measure the electrical current $I$ with a low noise current
to voltage amplifier as a function of source-drain \mbox{($V_{sd}$)} and gate \mbox{($V_g$)}
voltage and determine the differential conductance  \mbox{$G_d:=dI/dV_{sd}$}
numerically. Finally, the collected data $G_d(V_{sd},V_g)$
is represented in a two-dimensional grey-scale representation in which the grey-scale
corresponds to the magnitude of $G_d$. The linear-response conductance
$G:=I/V_{sd}$ with $V_{sd} \rightarrow 0$ is measured at a small but finite
source-drain voltage of \mbox{$40$\,$\mu$V}.

\section{Observation of Fano resonances in SWNTs}

Fig.~\ref{FanoFig2}a shows the differential conductance in the
form of a grey-scale plot of a single nanotube device for a
relatively large gate voltage range. The corresponding dependence
of the linear response conductance on the gate voltage $V_g$ is
displayed in Fig.~\ref{FanoFig2}b. This measurement is remarkable
in that the tunneling coupling between the nanotube and the
contacts must vary substantial, from weak coupling at large gate
voltage to very strong coupling at low gate voltage. For large
positive gate voltages (\mbox{$V_g \gtrsim 4$\,V}) a clear Coulomb
blockade pattern of low-conductance regions is observed which is
characteristic for weak coupling. As the gate voltage is reduced
higher-order tunnelling processes (so called cotunneling) start to
dominant, forming high conductance Kondo resonances around
zero source-drain voltage, \cite{Goldhaber-Gordon-1998,Kouwenhoven-Glazman-2001,Nygard-2000}
within the well known four-fold shell pattern generic for
high-quality carbon
nanotubes.\cite{Buitelaar-2002,Liang-2002,us-future} Reducing the
gate voltage even further and hence increasing the transparency to
the contact blurs the Coulomb blockade diamonds. This signals the
transition from a quantum dot to an open wire, which occurs if the
life-time broadening $\Gamma$ approaches the level spacing $\delta
E$, which in this device is of the same order as the charging
energy $U_C \sim 5$\,meV. In the limit $\Gamma >> U_C$,
interaction can be neglected. The overall transparency is then
expected to approaches unity, which for a single nanotube with
four channels relates to an upper bound in conductance of
$G=4e^2/h$. Because of residual backscattering at the contacts, a
weak periodic conductance modulation with gate voltage is
expected. This has been observed recently in SWNTs at a mean
transparency of $T=0.7$ and was termed the Fabry-Perot
interference effect.\cite{FP} The mean net transparency in our
device approaches $T=0.5$ for the lowest gate voltage.

Instead of a smooth continuation from the cotunneling to the
Fabry-Perot regime, sharp resonances appear below
\mbox{$V_g=2$\,V}. The two resonances visible in
Fig.~\ref{FanoFig2}a (labelled 1 and 2) are identified by their
asymmetric line-shapes as Fano resonances. To show this, the
measured gate dependence of the linear conductance for the two
resonances is magnified in Fig.~\ref{FanoFig3} (symbols) and shown
together with fits (solid curves) of Eq.~\ref{Fano1} using the
appropriate factor to convert gate voltage into energy (i.e.
$eC_g/C_{\Sigma}=0.025 \pm 0.05$). Here, we compare two parameters
which are obtained from the fitting procedure, \mbox{$\Gamma$} and
\mbox{$q$}. $\Gamma$ equals \mbox{$\approx 0.25$\,meV} and
\mbox{$\approx 0.5$\,meV} for resonance 1 and 2, respectively. For
comparison we have also extracted $\Gamma$ in the cotunneling
regime and found values \mbox{$\sim 2$\,meV}. Hence, the width of
the two new features is substantially smaller than the width of
the broadened nanotube levels. This difference gets even larger if
we take into account that $\Gamma$ grows further if one proceed
from the cotunneling regime at \mbox{$V_g\sim 3.5$\,V} to the
`open' regime at \mbox{$V_g\sim 2$\,V}. Both fits yield an
asymmetry parameter $q$ close to unity, i.e. \mbox{$q=-1$} and
\mbox{$q=-0.65$} for resonance 1 and 2, respectively. An asymmetry
parameter with a magnitude close to $1$ corresponds to asymmetric
line-shapes that are characteristic for Fano resonances, see
Fig.~\ref{FanoFig1}b. We note that both resonances have comparable
$q$ parameters of order $1$ and that the change in conductance is
for both cases large and of order $e^2/h$. The Fano fit is very
good for resonance 2 and it is reasonable for resonance 1. In the
latter case the deviations are getting appreciable away from the
resonance. Referring to the grey-scale plot in
Fig.~\ref{FanoFig2}a we see that this resonance is superimposed on
a low-conductance and strongly blurred Coulomb blockade diamond.
The assumption of the Fano description that the background
contributes to the interference with a constant non-energy
dependent term is only approximately valid here. These two
resonances will be analyzed further below.

The emergence of Fano resonances in single-wall carbon nanotubes
is exciting and we have therefore measured this sample again, now
lowering the gate voltage even further. This measurement is shown
in Fig.~\ref{FanoFig4}. A complex pattern of resonances appears
(arrows). We find resonances, anti-resonances and asymmetric Fano
lines shapes. All features resemble Fano resonances for different
$q$ values. Although the overall pattern looks quite irregular at
first sight, regular structures can be identified: In the first
place, one can identify `inverted' Coulomb blockade diamonds
(indicated with white arrows). Secondly, all resonances have
slopes which are quite comparable to the one in the Coulomb
blockade regime suggesting that the nanotube itself is the source
of the resonant state. The latter is also suggested by the fact
that in the whole gate voltage range the differential conductance
reaches the maximum conductance for a single SWNT of \mbox{$4
e^2/h$}, but never exceeds it.

\section{Discussion and Modelling}

The origin of the interfering paths, necessary for the Fano
effect, in a single quantum dot is often unclear.\cite{Gores,Zacharia}
One usually refers to the following picture: the non-resonant channel
can be seen as the direct transmission between the open source and open drain contacts, while
the weakly coupled resonant state corresponds to a bound state,
localized in the interior of the quantum dot.\cite{Clerk}

In carbon nanotubes the first account of Fano resonances have been
reported for crossed MWNTs.\cite{Kim} Because these authors have not observed
similar resonances in single tubes, they assigned the origin of the Fano resonance to
the particular geometry of two crossed tubes. In further accounts of Fano resonances
in MWNTs, the origin has been assigned to either an additional
carbon nanotube~\cite{Yi} acting as the non-resonant background or
to defects in the nanotube.\cite{Chandrasekhar}

Our experiment is the first for single-walled carbon nanotubes
(SWNTs) and the origin is puzzling as well. We point out, that an
individual SWNT contains two transport channels (not counting spin), leading
to the four-fold shell pattern of CNT quantum dots.\cite{Buitelaar-2002,Liang-2002,us-future}
In principle, this is enough for interference to occur. Assume that the two nearly
degenerate eigenstates are coupled with different strength to the
source and drain contacts, one with a large coupling and the other
with a weak one. Then, there is a broad and a narrow channel that
can interfere and give rise to Fano resonances. This is
schematically shown in Fig.~\ref{FanoFig1}c, where SWNT$_1$ and
SWNT$_2$ refer to the two orbital channels. However, we emphasize
that this picture is too simple for the following reason: The
appearance of Kondo resonances with similar Kondo temperatures at
intermediate gate voltage for each orbital state proves that they
are both coupled strongly to the contacts. It is quite unlikely
that the life-time broadening of one state increases while the
other decreases by lowering the gate voltage. Hence, we have to
explain the emergence of sharp Fano resonances based on two
(nearly) degenerate states which are both strongly coupled to the
contacts and have therefore both a large life-time broadening.

This problem has recently been studied theoretically in the limit
of vanishing interaction.\cite{Orelanna} Two quantum dot states
are coupled to the reservoirs with varying coupling parameters.
The calculation shows that even for {\em similar} coupling
strengths the two-dot ground state consists of a narrow and a wide
orbital. This is the result of hybridization leading to a
symmetric bonding and an asymmetric anti-bonding state. The latter
has a node at the contacts resulting in weak coupling to the leads
and therefore a small effective width $\Gamma$. If all coupling
terms were exactly the same, the life time of the anti-bonding
states would become infinite large. This never happens in practice, so
that one can expect intrinsic Fano resonances in SWNTs. This FR
was termed `ghost' FR by Ladr{\'o}n {\it et al.}\cite{Orelanna}
This is certainly the most attractive scenario. However, it is
obvious that Fano resonances can also occur `extrinsically' for
two separate individual SWNTs, provided they are geometrically
located within the phase-coherence length. Otherwise, a pure
superposition of two individual conductance patterns are expected
and not an interference effect. The problem of the interference
between two quantum dot states has recently also been addressed
using the scattering-(S)-matrix approach.~\cite{Kubala,Kubala1}
These authors also derive the correspondence between the Green's
function and S-matrix approach for this problem.

In the case of intrinsic Fano resonances (FRs) the regular
periodic pattern, which is evident in our measurement in the
Coulomb blockade and cotunneling regime, see Fig.~\ref{FanoFig3}a,
should evolve into a periodic pattern of FRs at higher tunneling
coupling. Although this looks promising in Fig.~\ref{FanoFig3}a,
the FR pattern in Fig.~\ref{FanoFig4}a is not as periodic as one
might expect. This is evidence in favor of extrinsic FRs. On the
other hand, the measured conductance is always bound by $4e^2/h$,
which is the maximum possible for a single tube. If two different
tubes interfere one would expect conductance values exceeding
$4e^2/h$ for certain gate voltages, which is not observed. This is
evidence in favor of intrinsic FRs. In the following we analyze
the observed FRs in greater detail.

Because we have measured both the gate and bias dependence we can
extend our analysis of the Fano resonances further and try to fit the
differential conductance in the vicinity of the resonance. Below, we
will do this for the two resonances 1 and 2. We model the problem
as two interfering channels in the Landauer-B\"uttiker formalism.\cite{Imry}
The transmission amplitude through the resonant channel is described
by \mbox{$t_r=\sqrt{T_r}i/(\epsilon+i)$}, where \mbox{$\epsilon=2(E-E_0)/\Gamma$}.
The square modulus of this function corresponds to a simple Lorenzian.
The transmission amplitude for the non-resonant channel is a constant
\mbox{$t_n=\sqrt{T_n}e^{i\phi}$}, where the phase \mbox{$\phi$} has
been introduced. Assuming spin degeneracy, the conductance at zero temperature
is given by \mbox{$G(\epsilon)=2e^2/h \cdot T_t(\epsilon)$} with the total
transmission probability \mbox{$T_t={\mid t_r+t_n \mid}^2$}. We obtain for $T_t$
\begin{equation}\label{eq:Fano2}
T_t=T_n+\frac{1}{1+\epsilon^2}\Big\{T_r+2\sqrt{T_rT_n}(\cos(\phi)+\epsilon
\sin(\phi))\Big\}.
\end{equation}
The differential conductance can then be obtained from
\begin{equation}\label{eq:Fano3}
\frac{\partial I}{\partial V_{sd}}=e^2/h\Big\{T_t(eV_{sd}/2-\alpha
V_g)+T_t(-eV_{sd}/2-\alpha V_g)\Big\},
\end{equation}
where $\alpha:=C_g/C_{\Sigma}$ denotes the gate-coupling strength as before.

We first discuss the symmetry of Eq.~\ref{eq:Fano3} and the
fitting procedure. The transmission amplitudes through the
resonant and non-resonant channel are chosen such that the linear
conductance is a Lorenzian for $\phi=0$, whereas it has a Fano
line-shape ($q=\pm 1)$ for \mbox{$\phi=\pm\pi/2$}. Far from the
resonance (\mbox{$\epsilon\rightarrow\infty$}) the conductance
asymptotically approaches the value of \mbox{$2e^2/h \cdot T_n$}.
For \mbox{$\phi=\pm\pi/2$}, the total transmission probability is
\mbox{$T_t$=$(T_n+T_r)$} at resonance. Note, that the
\mbox{$dI/dV_{sd}$} vs. \mbox{$V_{sd}$} characteristic will not
have mirror symmetry in general
for positive and negative gate voltages measured relative to the
center of the resonance.

The following fitting procedure has been adopted: First, we fit the
gate dependence of the linear conductance by changing the phase.
Then, the \mbox{$dI/dV_{sd}$} curves vs $V_{sd}$ are fitted for
specific gate voltages close to the resonance. The phase is
fixed, whereas \mbox{$T_n$, $T_r$, $\alpha$ and $\Gamma$} are free
parameters. Thereafter, the gate dependance is plotted again using
the average parameters from the fits found from the non-linear
regime. These average parameters are used to calculate the grey-scale plots.

For both resonances this procedure yields a phase close to $\pi/2$
which agrees reasonably with the asymmetry parameter of
\mbox{$q\approx -1$} deduced before. The fitting has been
preformed for four Fano resonances indicated with white arrows in
Fig.~\ref{FanoFig4}a, but we present only the results for
resonances 1 and 2. We start with resonance 2, which is a
particularly nice example. Fig.~\ref{FanoFig5}a shows the gate
dependence and Fig.~\ref{FanoFig5}b the bias dependence for four
different gate voltages. The solid curve in Fig.~\ref{FanoFig5}a
corresponds to the fit obtained from the linear conductance,
whereas the dashed one shows the resonance using the average
parameters deduced from the bias-dependence. We find for $\Gamma$
values of \mbox{$0.25-0.3$\,meV} in good agreement to the values
deduced before and a consistent gate-coupling parameter of
\mbox{$\alpha=0.02$}. The different values for \mbox{$T_n$ and
$T_r$} are plotted in Fig.~\ref{FanoFig5}c. The spread of
\mbox{$\approx \pm 0.15$} can be seen as a measure of the accuracy
of this procedure. Up to this error $T_n=1.3$ and $T_r=0.3$. The
average parameters are used to calculate the $dI/dV_{sd}$
grey-scale plot, which is shown together with the measurement in
Fig.~\ref{FanoFig5}d and e. A reasonable agreement is found. The
model clearly captures the main features and accounts for the
correct energy scales. The transmission probability of the
resonant channel of $T_r=0.3$ relates to a conductance of
\mbox{$0.6$\,$e^2/h$}, which is quite a substantial value.

We now turn to resonance 1. As we have pointed out before, the
agreement is less good here. This is due to the underlying blurred
Coulomb blockade structure, which results in a sizable suppression
of the conductance on one side of the resonance (left side), see
Fig.~\ref{FanoFig4}a. The result of the same procedure that led to
Fig.~\ref{FanoFig5} is shown in Fig.~\ref{FanoFig6} for resonance 1.
Fig.~\ref{FanoFig6}a displays the gate
dependence of the linear conductance. The solid curve corresponds
to the fit of the linear conductance vs gate voltage, whereas the
dashed curve has been calculated from the average parameters
deduced from $dI/dV_{sd}$ vs. $V_{sd}$. Because of the mentioned
suppression for \mbox{$\Delta V_g < 0$}, the differential
conductance has only been fitted in the non-linear regime for
three positive $\Delta V_g$. This is shown in
Fig.~\ref{FanoFig6}b, where we see that the fits match (apart from
the asymmetry) the measurements quite well. Due to the strongly
varying background a sizeable disagreement appears between the two
fits in Fig.~\ref{FanoFig6}a. However, we think that this can be
fully accounted for, by the background. The obtained parameters
$\phi$, $\alpha$ and $\Gamma$ compare very well with the ones
deduced before in  Fig.~\ref{FanoFig3}b: \mbox{$\phi=\pi/2$},
\mbox{$\alpha=0.02$}, and \mbox{$\Gamma=0.2$\,meV}. Not
surprisingly the two fitting procedures yield somewhat different
values for $T_r$ and $T_n$. $T_n \approx 1$ and $T_r \approx 1$
in one case and $T_n \approx 0.6$ and $T_r \approx 0.25$ in the
other.

Important for the following is the observation of an excitation
line which appears at negative bias voltage in the measurement and
is visible in Fig.~\ref{FanoFig6}b (arrows), as well as in the
grey-scale plot (inset). If we stick to the assumption that the
cause of the Fano resonances is intrinsic, the deduced excitation
energy of \mbox{$\delta E=0.6\pm0.1$\,meV} should then correspond
to the level spacing of this nanotube. We immediately see however,
that this value is much smaller than the measured level spacing of
\mbox{$\delta E \approx 5$\,meV} in this tube, which agrees very
well with the particle in a box model assuming spin and orbital
degeneracy for a nanotube of length \mbox{$300$\,nm},
corresponding to the contact separation. If on the other hand the
FR is extrinsic the resonant channel should correspond to a
nanotube with an effective length of as much as
\mbox{$L=2.8$\,$\mu$m}. This appears to be impossible, since the
electrodes are spaced by only \mbox{$300$\,nm}. However, already
in early work on carbon nanotubes (CNTs) two types of
characteristics have been found:\cite{Bockrath1} Contacting tubes
by evaporating metals over the tubes yielded `end-contacted'
CNTs,\cite{Bockrath2} whereas CNTs lying on metal electrodes
usually displayed a weaker coupling to the contacts and yielded
`bulk-contacted' tubes.\cite{McEuen,Tans3} Moreover, the single
electron level spacing $\delta E$ was found to agree with the
contact separation from edge to edge in the first case, whereas
the whole CNTs appeared to contribute, as apparent from small
values of $\delta E$ in the latter case. The states leading to
Fano resonances in our measurements are also much weaker coupled
to the leads. This is seen in the relative small $\Gamma$ values
deduced from the Fano resonances. Hence, the underlying resonant
channel may very well be a weakly coupled SWNT which resides in
one and the same bundle. This model may be regarded a likely
scenario because bundling in nanotubes is an ubiquitous
phenomenon. It is strong in arc-discharge and laser-evaporated
tubes, but it also occurs in CVD-grown CNTs which are considered
here.\cite{BabicKirchberg2004,Babic-Vibrating} We stress however,
that there is not prove that the excitation line corresponds to
the level spacing, which we have assumed before. There are also
electronic excitation with lower energies possible in carbon
nanotubes. For example, at half-filling (one electron on each
orbital) the exchange energy and level mismatch yield smaller
energy scales\cite{us-future}. Moreover, as we have emphasized
before already, if two tubes would contribute to transport, the
maximum conductance is not expected to be bound to $4e^2/h$.

We next compare the gate-coupling parameter for the Fano resonance
(FR) and the cotunneling regime. For the former (Fano resonances
labelled 1 and 2)  we have obtained \mbox{$\alpha= 0.02\pm
0.005$}, whereas \mbox{$\alpha=0.08\pm 0.01$} for the latter. This
is a significant difference amounting to a factor of four. This
difference is in favor of two tubes, one in effect short and the
other long as we explain now. Assume that there are indeed two
tubes contributing to the conductance in a small bundle. The
gate-capacitance $C_g$ can be assumed to be roughly equal, while
the capacitances to the leads should be strongly different. The
weakly coupled tube, which electrically appears to be much longer
than the contact separation, should have much larger source and
drain capacitances. The four-times smaller $\alpha$ relates into a
four-times larger total capacitance, and hence, into a four-times
smaller charging energy. The cotunneling regime of the dominant
tube yields \mbox{$U_C=5.3$\,meV}, so that the weakly coupled tube
should have a charging energy of  \mbox{$U_C\approx 1.3$\,meV}.
Together with the level spacing of \mbox{$\delta E \approx
0.6$\,meV} (assuming that this value does correspond to the level
spacing) yields an addition energy of \mbox{$\Delta E \approx
1.9$\,meV}. This relatively small addition energy may explain the
structure of the Fano resonances at \mbox{$V_g\approx 1.2,
0.8$\,V} which are shaped in a diamond-like pattern with an energy
scale corresponding to the reduced addition energy, see
Fig.~\ref{FanoFig4}a. However, at even smaller gate voltage of
\mbox{$V_g\approx 0.5$\,V} another `Fano-diamond' appears with
obviously a larger addition energy and with a gate-coupling
parameter that agrees with the cotunneling regime. Hence, the
interfering tube either evolves with increasing tunneling coupling
from a weakly coupled `long' tube to a stronger coupled
end-contacted one, or part of the observed Fano resonances are
intrinsic.

With regard to the phases, which were obtained by fitting resonance 1 and 2,
we mention that there is nothing peculiar about
the value of \mbox{$\phi=\pi/2$}.
The same fitting procedure has also been performed
for Fano resonances (FRs) labelled in Fig.~\ref{FanoFig4}a with \mbox{$3$ and $4$}
(not shown). Here we obtain the following parameters:
$T_n=0.8$, $T_r=0.2$, and  \mbox{$\Gamma=0.23$\,meV} for FR 3 and
$T_n=0.73$, $T_r=0.2$, and \mbox{$\Gamma=0.23$\,meV} for FR 4, while
the phase is now negative amounting to $\phi=-\pi/2$. $\alpha=0.02$ is consistent
with the previous value and the same for
both resonances. In addition, if we go
further out to even smaller gate voltages other phase values appear.

Finally, we briefly address the evolution of the Fano resonances
(FRs) at larger source-drain voltage $V_{sd}$. We observe that
most of the resonances vanish at \mbox{$|V_{sd}|\gtrsim
2.5$\,meV}. This can easily be understood by noting that
additional transport channels open up if \mbox{$V_{sd} > \delta
E$}. In particular if \mbox{$V_{sd} > \Delta E\approx 2$\,meV} the
resonant channel can involve excitations even for different charge
states. Because different phases are likely, the Fano resonance is
smeared out. A peculiar FR is observed at \mbox{$V_g\approx
1.3$\,V} in Fig.~\ref{FanoFig4}a (dashed arrow). This resonance
starts with a large gate-coupling $\alpha$ around zero
source-drain voltage $V_{sd}$ and evolves into a smaller coupling
parameter $\alpha$ for larger $V_{sd}$. This apparent curving,
which can also take up the reversed order, is currently not
understood.

\section{Summary}

We have observed Fano resonances in CVD-grown SWNTs. The measured
conductance evolves with reducing gate voltage from the
single-electron tunneling regime at low coupling to the
cotunneling regime at intermediate coupling. If the transparency
to the contacts are increased further sharp resonances emerge
superimposed on a weakly varying background. These resonances are
identified as Fano resonances and can reasonably well be modelled
with the traditional Fano approach of interference between a
resonant and a non-resonant channel without interaction. The source
of the two channels may reside in different tubes within one
bundle, one weakly coupled to the contacts and the other strongly.
This scenario is appealing because bundling is an ubiquitous
phenomenon in the growth process of carbon nanotubes. It also
occurs in CVD-grown tubes, which are studied here. However, it is
quite surprising that the measured electrical conductance is bound
by $4e^2/h$, which is the maximum possible for a {\em single}
nanotube. In case of the interference between two and more
nanotubes one would expect the conductance to exceed this value
for certain gate voltage values. We have pointed out, that Fano
resonances may also arise intrinsically, even if the two orbital
states are both strongly coupled to the contacts. Future work must
clarify the role of intrinsic and extrinsic Fano resonances in
SWNTs.

\begin{acknowledgments}
We acknowledge fruitful discussions with W. Belzig, T. Kontos and
D. Loss. This work has been supported by COST and the RTN DIENOW (EU-5th framework),
the Swiss NFS, and the NCCR on Nanoscience.
\end{acknowledgments}


\begin{figure}[htb!]
\begin{center}
\end{center}
\caption{\label{FanoFig1} (a) Schematic view of a Fano system
consists of a resonant channel through, e.g. a quantum dot (QD),
and a non-resonant channels. (b) Normalized Fano line-shapes
calculated from Eq.~\ref{Fano1} for several asymmetry \mbox{$q$}
parameters. (c) A scheme of two SWNTs connected to the left (L)
and right (R) leads. Due to different coupling strength the
zero-dimensional states of each SWNT acquires a different width
expressed by $\Gamma_{1,2}$. The two interfering channels may be
due to two `individual' SWNTs of a bundle or may represent the two
transport channels of one and the same SWNT.}
\end{figure}

\begin{figure}[htb!]
\begin{center}
\end{center}
\caption{\label{FanoFig2} (a) Grey-scale representation of the
differential conductance (\mbox{$dI/dV_{sd}$}) versus bias
(\mbox{$V_{sd}$}) and gate voltage (\mbox{$V_g$}), and (b), the
corresponding linear response conductance versus gate voltage. Due
to the strong dependence of the tunneling coupling to the leads on
gate voltage, several physical phenomena are observed together.
These are from right to left: Coulomb blockade, the Kondo effect
and Fano resonances, corresponding to regimes of low, intermediate
and high tunneling coupling. Note, that the conductance
dramatically increases for \mbox{$V_g\lesssim3$\,V} reaching a
maximum value of \mbox{$G\approx 4e^2/h$} as expected for an ideal
metallic SWNT.}
\end{figure}

\begin{figure}[htb!]
\begin{center}
\end{center}
\caption{\label{FanoFig3} Comparison of the measured (symbols)
linear response conductance $G$ for the resonances labelled 2 (a)
and 1 (b) in Fig.~\ref{FanoFig2}a with the Fano formula
Eq.~\ref{Fano1} (solid curves). $\Delta V_g:= V_g - V_g^0$, where
$V_g^0$ denotes the gate voltage at the center of the resonance:
$V_g^0=1.9$, \mbox{$1.54$\,V} for resonance 1 and 2. The extracted
values from the fits to the Fano equation are: $q=-1.0\pm 0.16$,
$-0.63 \pm 0.05$ and $\Gamma = 0.25\pm 0.05$, \mbox{$0.49 \pm
0.08$\,meV} for resonance 1 and 2, respectively.}
\end{figure}

\begin{figure}[htb!]
\begin{center}
\end{center}
\caption{\label{FanoFig4} (a) Differential conductance
(\mbox{$dI/dV_{sd}$}) versus bias voltage (\mbox{$V_{sd}$}) and
gate voltage (\mbox{$V_g$}). Dark correspond to high
(\mbox{maximum$=4e^2/h$}) and white to low conductance. Fano
resonances are indicated with the arrows. (b) Corresponding linear
response conductance $G$.}
\end{figure}

\begin{figure}[htb!]
\begin{center}
\end{center}
\caption{\label{FanoFig5} Comparison between the measured
differential conductance of the Fano resonance 2 (see
Fig.~\ref{FanoFig4}a) and fits to Eq.~\ref{eq:Fano3}. In (a) the
dashed curve represents the best fit taking into account only the
linear response conductance, whereas the solid curve is calculated
from average parameters deduced by fitting the non-linear
differential conductance  $dI/dV_{sd}$ vs. $V_{sd}$ for a set of
gate voltages, shown in (b). The curves are vertically offset by
$e^2/h$ for clarity. For a) the whole set of parameters are given
in the figure, whereas we only represent the deduced transmission
probabilities for the non-resonant $T_n$ (open circle) and
resonant $T_r$ (full circle) channel of part (b) in (c). The
calculated differential conductance (e) is compared with the
measured one (d). The parameters are \mbox{$T_n=1.3$, $T_r=0.3$,
$\Gamma=0.25$\,meV, $\phi=\pi/2$} and \mbox{$\alpha=0.02$}.}
\end{figure}

\begin{figure}[htb!]
\begin{center}
\end{center}
\caption{\label{FanoFig6} Comparison between the measured
differential conductance of the Fano resonance 1 (see
Fig.~\ref{FanoFig4}a) and fits to Eq.~\ref{eq:Fano3}. In (a) the
solid curve represents the best fit taking into account only the
linear response conductance, whereas the dashed curve is
calculated from average parameters deduced by fitting the
non-linear differential conductance  $dI/dV_{sd}$ vs. $V_{sd}$ for
a set of (only) positive gate voltages, shown in (b). The curves
are vertically offset by $e^2/h$ for clarity. For a) the whole set
of parameters are given in the figure, whereas the parameters for
the three fits in (b) are: $\blacksquare$: $T_n=0.85$, $T_r=0.9$,
\mbox{$\Gamma=0.25$\,meV}, $\phi=\pi/2$, and $\alpha=0.027$;
$\square$: $T_n=1$, $T_r=1.2$, \mbox{$\Gamma=0.25$\,meV},
$\phi=\pi/2$, and $\alpha=0.02$; $\scriptsize{\bigcirc}$:
$T_n=1.1$, $T_r=1.4$, \mbox{$\Gamma=0.25$\,meV}, $\phi=\pi/2$, and
$\alpha=0.017$. The observed asymmetry in bias voltage originates
from different coupling of the resonant states to the leads. The
inset in (b) shows the measured differential conductance. An
excited state is visible for $V_{sd}<0$, both in the greyscale
plot and the $dI/dV_{sd}(V_{sd})$ (arrows).}
\end{figure}


\begin{thebibliography}{99}
\bibitem{us-future}
A detailed analysis of the Kondo effect, in particular the
fundamental question on the nature of the ground state at
half-filling will be published in a different paper.
\end{thebibliography}
\end{document}